\documentclass[pre,aps,reprint,twocolumn,footinbib,longbibliography,floatfix]{revtex4-2}
\usepackage{graphicx}
\usepackage{dcolumn}
\usepackage{bm}
\usepackage{amssymb}
\usepackage{amsfonts,mathdots}
\usepackage{pifont}
\usepackage{amsmath}
\usepackage{times}

\begin{document}

  \title{Discontinuous transition to chaos in a canonical random neural network}
  \author{Diego Paz\'o}
\affiliation{Instituto de F\'{i}sica de Cantabria (IFCA), Universidad de
Cantabria-CSIC, 39005 Santander, Spain}
\date{\today}
 \begin{abstract}
We study a paradigmatic
random  recurrent neural network introduced by Sompolinsky, Crisanti, and Sommers (SCS).
In the infinite size limit, this system exhibits a direct
transition from a homogeneous rest state to chaotic behavior, with
the Lyapunov exponent gradually increasing from zero.
We generalize the SCS model considering odd saturating nonlinear
transfer functions, beyond the usual choice $\phi(x)=\tanh x$.
A discontinuous transition to chaos occurs
whenever the slope of $\phi$ at 0 is a local minimum (i.e.,~for $\phi'''(0)>0$).
Chaos appears out of the blue, by an attractor-repeller fold.
Accordingly, the Lyapunov exponent stays away from zero at the birth of chaos.
 \end{abstract}

\maketitle

\section{Introduction}

Random neural networks are interesting mathematical
abstractions for neuroscience \cite{GKN+14} and
several machine learning techniques \cite{jaeger04}.
In these systems the point of optimal computational
performance is generally expected to be at (or near) the 'edge of chaos'
\cite{packard88,langton90,edge,toyoizumi11,keup21}.
Characterizing the onset chaos in random neural networks
represents
therefore a significant scientific endeavor \cite{kadmon15,kusmierz20}.

A prototypic random neural network
was put forward long ago by Sompolinsky, Crisanti,
and Sommers (SCS) in Ref.~\cite{SCS88}. 
This system became a workhorse model
for mathematical neuroscience,
at the core of more sophisticated models
of plasticity and learning \cite{sussillo09,clark_abbott},
working memory \cite{barak13,pereira23},
multiplicative gating interactions \cite{schwab22}, etc.

The SCS model consists of
$N\gg1$ neurons (or `rate units'), whose activities $\{h_i\}_{i=1,\ldots,N}$
are scalar variables.
They are governed by $N$ first-order ordinary differential equations:
\begin{equation}
{\dot h}_i=-h_i + g\sum_{j=1}^N J_{ij} \phi(h_j) .
\label{scs}
\end{equation}
The coupling between them is weighted by the $N\times N$ matrix
$J$, whose off-diagonal elements are independently drawn at random
from a Gaussian distribution of zero mean and variance $1/N$.
In addition, self-coupling is excluded setting $J_{ii}=0$.
The so-called transfer, activation, or gain function $\phi$ possesses
a sigmoidal shape.
Finally, $g>0$ is the coupling constant (or `synaptic gain' in
jargon).

Our current comprehension of the SCS model within its basic setup,
$\phi(x)=\tanh x$ and large size $N$,
is impressive.
In the thermodynamic limit ($N\to\infty$),
a direct transition from
a stable fixed point at the origin (FP$_0$) to chaotic behavior occurs when $g$ is increased above $1$.
The level of
``chaoticity'' gradually increases above $g=1$: the asymptotic
growth of the largest Lyapunov exponent (LE)  is $\lambda\simeq(g-1)^2/2$ \cite{SCS88,CS18}.
At the same time, the emerging chaotic attractor is not low-dimensional:
its dimension grows linearly with the system size $N$---i.e., chaos is extensive \cite{engelken,clark23}.
Moreover, the onset of chaos coincides with the birth of an unstable heterogeneous rest state.
For finite $N$ this translates into
an explosion in the
number $n$ of (unstable) fixed points,
scaling exponentially with $N$: $n\sim e^{(g-1)^2 N}$
for $g$ just above $1$ \cite{wainrib13,stubenrauch23}.

The scenario just described is markedly different from
the classical ``low-dimensional'' routes to chaos,
and from the onset of phase turbulence or
spatio-temporal intermittency in extended systems \cite{Manneville}.
It is not completely robust though.
For instance, noise \cite{molgedey92,schuecker18}
or heterogeneous external inputs \cite{stubenrauch23} alter the transition to chaos as described above.
Still, in both cases the ``chaoticity'' (i.e., the LE) smoothly grows from zero.
In parallel, considerable effort has been devoted to generalize the
coupling matrix in \eqref{scs}
\cite{aljadeff15,marti18,mastrogiuseppe,ipsen20}. To our knowledge, 
these works do not contradict the expectation that
chaos sets in smoothly, with a  LE growing from zero above threshold.
Only recently, a discontinuous transition to chaos has been detected in
the SCS model supplemented with gating
interactions \cite{schwab22} and in a discrete-time neural connectivity model \cite{kusmierz20}.

In the context of the SCS model the influence of the
transfer function $\phi$ remains largely ignored, since it is
generally believed that adopting the hyperbolic tangent serves as a
``prototype of generic odd symmetric saturated sigmoid functions'' \cite{CS18}.
In this paper we make this assertion more quantitative.
We find that the sign of $\phi'''(0)$ determines if the
transition to chaos is qualitatively as described above,
or discontinuous. In the latter case
---occurring when $\phi'''(0)>0$---  stable FP$_0$ and a chaotic attractor
coexist in a finite interval of $g$.
The chaotic attractor appears out of the blue
in an attractor-repeller collision between chaotic sets.
Accordingly, the LE stays away from zero at the bifurcation point.

This paper is organized in the following way.
In Sec.~II we introduce a monoparametric family of transfer functions,
and show the results of our simulations, in particular the coexistence mentioned above.
Section III provides a theoretical explanation for the observed results, and
contains the main results of this work. In Sec.~IV 
we show a couple of additional simulations. Finally, Sec.~V serves to
recall the main conclusions of this work, and to provide some outlook.

\section{Preliminary Numerical results}

\begin{figure}
    \centering
    \includegraphics[width=0.8\linewidth]{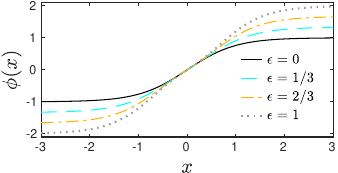}
    \caption{Transfer function $\phi(x)$ for different values of parameter $\epsilon$.}
\label{fig:phi}
\end{figure}

Throughout this work numerical simulations are carried out
with a family of odd symmetric transfer functions:
\begin{equation}
\phi(x)=\tanh x + \epsilon  \tanh^3 x .
\label{tf}
\end{equation}
For $\epsilon=0$  we recover the usual choice $\phi(x)=\tanh x$.
And the monotonicity of $\phi(x)$ is preserved for $\epsilon>-1/3$.
Moreover, the slope remains 1 at the origin, irrespective of the $\epsilon$ value.
Function $\phi(x)$, saturates at $\pm (1+\epsilon)$ as $x\to\pm\infty$.
For the sake of illustration, $\phi(x)$
is represented in Fig.~\ref{fig:phi}
for several $\epsilon$ values.

We start exploring  the dynamics of the model defined by Eqs.~\eqref{scs} and
\eqref{tf} numerically. We consider
one specific network realization of size $N=10^3$.
Linearizing the model, and computing the eigenvalues of the
corresponding $10^3\times10^3$ Jacobian matrix, we determined when
the trivial fixed point $h_i=0$ loses its stability. For the particular network
the critical coupling turned out to be $g_0=1.006\ldots$,
not far from $g_c=1$, the actual stability boundary of FP$_0$ 
in the thermodynamic limit.
Note that the value of $g_0$ is independent of $\epsilon$
since $\phi'(0)=1$ always holds.

We integrated Eq.~\eqref{scs} using a fourth order
Runge-Kutta algorithm with time step $\Delta t=10^{-2}$.
The state of the network was tracked measuring the
variance:
\begin{equation}
\Delta(t)=\overline{h_i^2} - \overline{h_i}^2 .
\end{equation}
The overline denotes the population average.
In Fig.~2(a), we show the results for six different
values of $\epsilon$. In each case the simulation started
at a high value of $g$, and then its value was decreased quasi-adiabatically.
For each value of $g$ (and $\epsilon$) the local maxima of
$\Delta(t)$ are represented by dots, while the time
average $\langle\Delta\rangle$ is identified by one square.
The LE $\lambda$ was also computed,
see Fig.~2(b).
At parameter values with positive LE, i.e. chaos, the
local maxima of $\Delta$ are scattered, as expected.
In contrast, when only a few
different local maxima of $\Delta$ are measured
the LE is approximately zero, indicating periodic dynamics.

\onecolumngrid

\begin{figure*}[b!]
    \centering
    \includegraphics[width=0.8\linewidth]{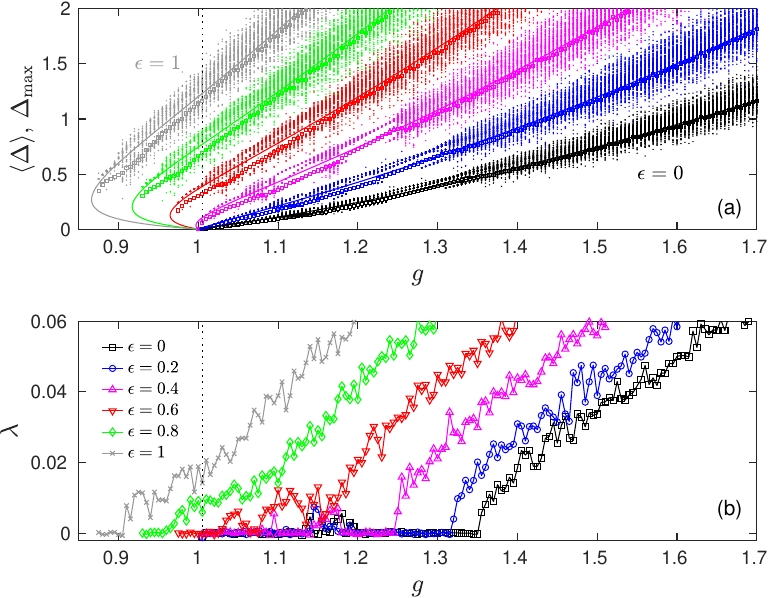}
    \caption{Numerical results for a particular network of $N=10^3$ units
    and six different values of $\epsilon=\{0,0.2,0.4,0.6,0.8,1\}$.
    The coupling constant $g$ was decreased at steps of size $\Delta g=5\cdot10^{-3}$.
     The simulations were finished when the system falls into the trivial fixed point FP$_0$.
    (a) The mean variance $\langle \Delta\rangle$ and
    local maxima of $\Delta(t)$---denoted by $\Delta_{\mathrm{max}}$---are depicted for each $g$ value
    with squares and dots, respectively.
The vertical dashed line signals the stability
    threshold of the homogeneous fixed point, located at $g_0\simeq1.006$. The trivial
    fixed point attractor $h_i=0$ present for $g<g_0$ is not shown.
    Solid lines are the variances $c_0$ of chaotic solutions
    in the thermodynamic limit, predicted from Eq.~\eqref{gcaos}.
    (b) Largest Lyapunov exponent of the attractors in panel (a).
    }
\label{fig:fig2}
\end{figure*}

\clearpage

\twocolumngrid

If we look at the results for the canonical case $\epsilon=0$,
it is apparent that
chaos ---signified by a positive LE--- disappears at a certain value of $g$ above $g_c$ (from above).
This is the kind of finite-size effect that
we must keep in mind. Finite networks do not exactly behave as the
theory, valid in the thermodynamic limit, dictates, specially
near the bifurcation points \cite{engelken,curato}.
There, the results between different realizations
of the coupling matrix will be specially conspicuous. This drawback
is minimized increasing the network size, but it is difficult in practice since
the computational cost grows quite rapidly with the system size.

Let us now discuss the results for $\epsilon>0$ in Fig.~2.
The results for $\epsilon=0.2$ are qualitatively similar to those for $\epsilon=0$. 
The variance attains larger values in the former case because the transfer function
saturates at a larger values (in absolute value).
Qualitatively new results are found for larger
$\epsilon$: Variances remain above zero
for $g$ below $g_c$.
The most remarkable observation is the
bistability between chaos (or periodic dynamics)
and the trivial rest state (FP$_0$).
As occurred for small $\epsilon$, the nontrivial attractor
becomes periodic before disappearing.
One might argue that the finiteness of the network, or poorly
treated transient dynamics,
are responsible of the peculiarities observed.
As we show next, our analysis,
valid in the thermodynamic limit,
successfully explains the behavior of the system.

\section{Theory}

\subsection{Dynamic mean-field analysis}

Analyzing the model \eqref{scs}, even assuming $N\to\infty$,
is a difficult task.
Either a path integral approach \cite{CS18} or
dynamic mean-field theory \cite{SCS88,kadmon15,HD} can be used.
We do not review these mathematical treatments here.
Instead, we simply borrow the final results
as recently published in \cite{HD}, and apply them
to our transfer function \eqref{tf}.

In the original work by SCS \cite{SCS88}, it was found that the
autocovariance
\begin{equation}
c(\tau)= \langle h_i(t) h_i(t+\tau) \rangle
\label{ctau}
\end{equation}
is a key quantity.
Note that $c(\tau)$ in Eq.~\eqref{ctau} solely depends
on the time difference, as the dynamics is assumed to settle into a stationary state.
For $\tau=0$, we recover the variance $c(0)\equiv c_0$
($\langle h_i \rangle=0$ for $N\to\infty$). Eventually, we shall compare $c_0$ with the
time average $\langle\Delta\rangle$ computed in Fig.~2. Both quantities
should coincide in the thermodynamic limit as, due to symmetry, 
we expect $\lim_{N\to\infty} \overline{h_i}=0$.

One may can prove that
the autocovariance obeys a second-order ordinary differential equation\cite{SCS88,HD}:
\begin{equation}
\frac{d^2c(\tau)}{d\tau^2}=-V'(c;c_0) .
\label{cdd}
\end{equation}
The constant $c_0$ plays two roles in this equation:
it is a parameter of the potential $V$, and
it specifies the initial condition for $c_0=c(0)$.
The value of $c_0$ has to be determined requiring self-consistency.
The exact form of $V$ is \cite{HD}:
\begin{equation}
V(c;c_0)=-\frac{c^2}2 + g^2 f_\Phi(c,c_0)-g^2 f_\Phi(0,c_0) ,
\label{pot}
\end{equation}
with
\begin{eqnarray}
&&f_\Phi(c,c_0)=\\
&&\iint_{-\infty}^{\infty} \Phi\left(\sqrt{c_0-\frac{c^2}{c_0}}z_1
+ \frac{c}{\sqrt{c_0}}z_2\right) \Phi\left(\sqrt{c_0}z_2 \right) Dz_1 \, Dz_2.
\nonumber
\end{eqnarray}
Here,
$Dz$ is the Gaussian integration measure $Dz=\exp(-z^2/2)/\sqrt{2\pi} dz$,
and $\Phi$ is the integral of $\phi$:
\begin{equation}
\Phi(x)= \int_0^x \phi(x') dx'=(1+\epsilon) \ln(\cosh x)-\frac\epsilon2
\tanh^2x.
\end{equation}

The last term in Eq.~\eqref{pot} is an offset to ensure $V(0,c_0)=0$.
As already explained, the shape of the potential depends on
the constant $c_0$, acting as a parameter that must be
self-consistent with the dynamics. For instance,
a (heterogeneous) fixed point corresponds to
a $\tau$-independent covariance $c_*$, which is
satisfied by an equilibrium point of Eq.~\eqref{cdd}. In turn
$V(c;c_*)$ must exhibits an extremum precisely at $c=c_*$.
As a double-check of our numerical integral solvers,
we verified that the
variance $c_*$ of the heterogeneous fixed point,
obtained through static mean-field theory (see Appendix)
corresponded to a minimum of the potential of $V(c;c_*)$
exactly at $c=c_*$. Periodic orbits, corresponding to
periodic $c(\tau)$ are also possible,
but we do not consider them. It was already
concluded for $\epsilon=0$, that such orbits are unstable \cite{SCS88},
and there is not physical reason to expect them to become
stable for $\epsilon>0$.

\subsection{Chaotic solutions}

\begin{figure}
    \centering
    \includegraphics[width=0.8\linewidth]{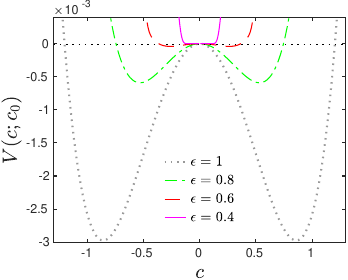}
    \caption{Potential $V(c;c_0)$ for $g=1$ and four  different values of $\epsilon$.
    In each case, the value of parameter $c_0$ was selected
    to correspond to the chaotic state,
    i.e.~a trajectory obeying Eq.~\eqref{cdd}, with initial condition $c(0)=c_0$,
    converges to $0$ as $\tau\to\infty$.
    The $c_0$ values are
    $c_0\simeq1.203$, $0.745$, $0.365$, and $0.0736$;
    in order of decreasing $\epsilon$.
    }
\label{fig:pot}
\end{figure}

For a chaotic solution, $c_0$ is such that,
imposing the initial condition $c(0)=c_0$ (and $\dot c(0)=0$ as noise
is absent \cite{HD}) in Eq.~\eqref{cdd},
the asymptotic behavior is
$\lim_{\tau\to\infty} c(\tau)=0$,
i.e.~the autocorrelation vanishes in the infinite-$\tau$ limit.
As the initial and the asymptotic points possess null potential and kinetic
energies, the self-consistent condition boils down to
\begin{equation}
V(c_0;c_0)=0 .
\label{0}
\end{equation}
Needless to say, this condition can be fulfilled for certain values of
$\epsilon$ and $g$, but not for others.
For example, at the critical value $g=1$, when $\epsilon$ is small or zero
the only self-consistent solution is $c_0=0$ (i.e., the fixed point FP$_0$) .
Contrastingly, for large enough $\epsilon$, 
other self-consistent potentials exist.
Four of them are shown in Fig.~\ref{fig:pot}.

In order to locate chaotic solutions in parameter space,
it is convenient to minimize the numerical effort.
Imposing the condition in Eq.~\eqref{0} to Eq.~\eqref{pot} we obtain
$g$ as a parametric function of $c_0$: 
\begin{eqnarray}
g^2_{\mathrm{ch}}(c_0)&=&\frac{c_0^2/2}{f_\Phi(c_0,c_0)-f_\Phi(0,c_0)} \label{gcaos}\\
&=&\frac{c_0^2/2}{\int [\Phi(\sqrt{c_0} z)]^2 Dz- [\int\Phi(\sqrt{c_0} z) Dz)]^2} .\nonumber
\end{eqnarray}
The integrals in this expression were solved using {\sc mathematica}, and
the results are depicted as solid lines in Fig.~\ref{fig:fig2}(a).
For large enough $\epsilon$ two chaotic solutions coexist
in a range of $g$ values below 1.
As one could have presumed, the numerical simulations indicate that the branch
with the highest variance corresponds to an attractor, while the lower branch
corresponds to a chaotic saddle, see Sec.~\ref{sec:saddle} below.

\begin{figure}[b]
    \centering
    \includegraphics[width=0.8\linewidth]{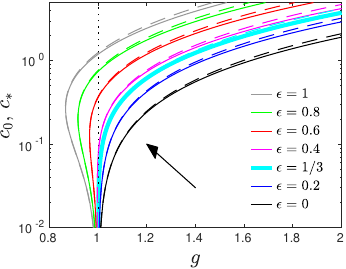}
    \caption{Solid line: Variance $c_0$ of the chaotic solutions,
    obtained from Eq.~\eqref{gcaos}.
    Dashed line: Variance $c_*$ of the heterogeneous rest state,
    obtained from Eq.~\eqref{selfc}.
    The arrow indicates the direction of growing $\epsilon$.
    The curves corresponding to $\epsilon>1/3$ exhibit a fold.
    }
\label{fig:fold}
\end{figure}

\subsection{Critical $\epsilon$}

We elucidate next what determines if the line of chaos emerging at $g=1$
has positive or negative slope.
We start assuming $0<c_0\ll1$, and Taylor expand $\Phi(\sqrt{c_0}z)$ in Eq.~\eqref{gcaos}.
Taking into account that $\phi$ is an odd symmetric function, and $\phi'(0)=1$, we get
\begin{equation}
g_{\mathrm{ch}}^2\simeq\frac1{1+\phi'''(0) c_0+O(c_0^2)} \label{gcaosb} .
\end{equation}
In this formula the sign of the third derivative of $\phi$ at zero
determines the existence of the solution either above or below $g=1$.
The usual scenario is observed for negative $\phi'''(0)$, i.e.,
when the slope of $\phi$ at the origin is a local maximum. Otherwise, if $\phi'''(0)>0$,
then the chaotic branch emanates with negative slope, and
a fold develops.

\begin{figure}[b]
    \centering
    \includegraphics[width=0.8\linewidth]{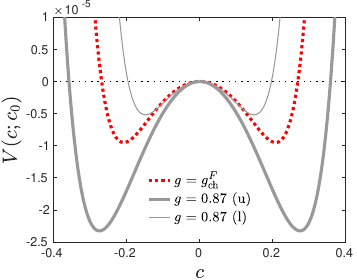}
    \caption{Self-consistent potentials $V(c;c_0)$ for $\epsilon=1$ and two different values of
    the coupling constant $g$.
    Dashed red line: The single potential at the fold,
    $g_{\mathrm{ch}}^F=0.866216\ldots$.
    {Solid} gray lines: The pair of self-consistent potentials at $g=0.87$ (slightly above $g_{\mathrm{ch}}^F$).
    The thick line corresponds to the upper branch (u),
    and the thin line to the lower branch (l).
    The values of the variances are $c_0\simeq0.1964$, $0.269$, and $0.358$.
    }
\label{fig:pot_eps1}
\end{figure}

For our particular transfer function in Eq.~\eqref{tf}, $\phi'''(0)=-2+6\epsilon$, and
therefore the critical $\epsilon$ value turns out to be $\epsilon_c=1/3$.
In Fig.~\ref{fig:fold} we illustrate the effect of crossing this value,
displaying $c_0$ as function of $g$ (in logarithmic scale).
For the crossover value $\epsilon=1/3$ a {thick} cyan line is displayed.
A fold is apparent
for all $\epsilon$ values above $1/3$.
We interpret this fold as an attractor-repeller collision of chaotic sets,
with the upper (lower) branch corresponding to the attractor (saddle).
Numerical simulations do not contradict this expectation.
For illustration, two potentials
corresponding to the upper and lower branches of chaotic
dynamics are shown by {solid} grey lines in Fig.~\ref{fig:pot_eps1}.

As mentioned in the Introduction, in the SCS model (with $\epsilon=0$)
an unstable heterogeneous fixed point bifurcates from FP$_0$ at $g=1$.
In the Appendix we determine the
variance of the heterogeneous fixed point $c_*$ using static mean-field theory.
In Fig.~\ref{fig:fold} the variance $c_*$
is depicted by a dashed line for the same values of $\epsilon$ that $c_0$.
Both quantities exhibit a fold for $\epsilon>1/3$.
This means that varying $g$ the system
exhibits a fold (or saddle-node) bifurcation of fixed points
(in infinitely many dimensions).
The pair of fixed points are both unstable.
This is analytically investigated  in the Appendix.
We can also appreciate in Fig.~\ref{fig:fold}
that the values of $c_*$ and $c_0$
are asymptotically the same in the $g\to1$ limit,
cf.~Eq.~\eqref{gcaosb} and Eq.~\eqref{g2}
in the Appendix.

\subsection{Finiteness of the Lyapunov exponent at the fold}

As already explained, if $\epsilon>1/3$, then
chaos appears out of the blue
at certain $g_{\mathrm{ch}}^F$ in an attractor-repeller collision
(assuming the $N\to\infty$ limit).
Simulations indicate that only the upper branch of the chaotic solution
is attracting. The lower branch, corresponding  to a repeller,
terminates at $g=1$.
In Fig.~\ref{fig:pot_eps1} the self-consistent potential at the
critical value $g_{\mathrm{ch}}^F$ for $\epsilon=1$
is depicted by a red dashed line.
This potential does not exhibit any peculiarity
(in contrast with the normal scenario when $g\to1^+$).
The  covariance $c(\tau)$ will approach zero at a finite decay rate.
This indicates the LE is necessarily above zero.

\begin{figure}[b]
    \centering
    \includegraphics[width=0.8\linewidth]{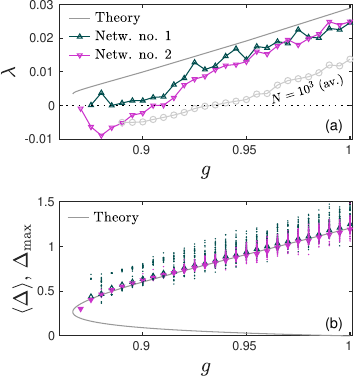}
    \caption{Behavior of two different networks of size $N=4\times10^3$
    as a function of $g$ ($\epsilon=1$).
    (a) The LE for
    each network.
    The theoretical value of
    the LE (upper branch) obtained from Eqs.~\eqref{schroedinger}
    and \eqref{le}, as well as
    the average LE over 20 network
    configurations with $N=10^3$, are
    {depicted by a solid line and circles, respectively.}
(b) Mean variance $\langle \Delta\rangle$ and
    maximum values of $\Delta(t)$ for the same attractors than in panel (a).
    The solid line is the variance $c_0$ predicted by dynamic
    mean-field theory, Eq.~\eqref{gcaos}.
    }
\label{fig:N4000}
\end{figure}

Moreover, as originally found \cite{SCS88}, the LE is related with
the quantum-mechanical ground-state energy
of a particle confined in a symmetric potential well
$W(\tau)=-V''(c(\tau))$.
Given the time-independent Schr\"odinger
equation:
\begin{equation}
-\psi''(\tau)+W(\tau) \psi(\tau)=E \psi(\tau) ,
\label{schroedinger}
\end{equation}
the energy of the ground state $E_0$ yields
the LE:
\begin{equation}
\lambda=-1+\sqrt{1-E_0} .
\label{le}
\end{equation}
In our numerical implementation, we started fitting $V(c)$
by an even polynomial, such that the nontrivial zero
coincided with $c_0$ up to the six decimal digit (at least).
The second derivative of the polynomial was evaluated at
numerical solutions of $c(\tau)$ obtaining in this way $W(\tau)$.
Finally, the value of $E_0$ was obtained applying the shooting method
to Eq.~\eqref{schroedinger}. We impose $\psi'(0)=0$ and
iteratively search a solution $\psi(\tau)$ monotonically decreasing to zero
(since the ground state is even symmetric and node-free).

The previous numerical scheme was applied to the upper chaotic branch
for $\epsilon=1$ and $g\le1$. The result is
shown by a {solid} grey line in Fig.~\ref{fig:N4000}(a).
The LE disappears with a finite value at $g=g_{\mathrm{ch}}^F$.
This is precisely what should occur in
an attractor-repeller collision between
chaotic solutions, see {the solid} grey line in Fig.~\ref{fig:N4000}(b).

As a complement to the theoretical results we also show in Fig.~\ref{fig:N4000} the
results for two network realizations with $N=4\times10^3$.
In Fig.~\ref{fig:N4000}(a) we additionally represent  the average over 20
realizations for $N=10^3$ (a few realizations with no chaos were discarded).
The numerical results are fully consistent with the theory.
The gap between theory and numerics
narrows as the system size increases.

It is important to keep in mind that
a finite network necessarily behaves as
a generic dynamical system (with global inversion symmetry $h_i\to-h_i$).
Hence, the disappearance of the nontrivial chaotic
state necessarily follows a standard route, instead of an
attractor-repeller collision of chaotic sets. However, the larger the network
the closer to the asymptotic scenario.

\subsection{Phase diagram}

\begin{figure}[b]
    \centering
    \includegraphics[width=0.8\linewidth]{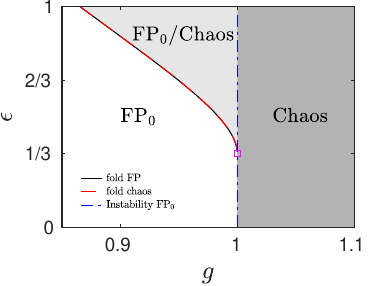}
    \caption{Phase diagram of the SCS model with the transfer function in Eq.~\eqref{tf}.
    In the light shaded
    region, the chaotic attractor and the stable trivial fixed point (FP$_0$) coexist.
    The pink square is located at the doubly degenerate point $(g,\epsilon)=(1,1/3)$.
    }
\label{fig:pd}
\end{figure}

We end this section condensing our previous results
in the phase diagram displayed in Fig.~\ref{fig:pd}.
The fold of the fixed point and the
fold of chaos almost overlap, cf.~Fig.~\ref{fig:fold}.
Increasing $g$, the pair of (unstable) fixed points appear
prior to the pair of chaotic states (attractor and repeller).
The distance between both fold bifurcations is tiny.
For example, at $\epsilon=1$ the folds for chaos and
the fixed point are at $g_{\mathrm{ch}}^F\simeq0.8662$ and $g_{\mathrm{fp}}^F\simeq
0.8655$, respectively.
For a large finite network,
this means that topological complexity (an exponential number of fixed points),
precedes dynamical complexity (chaos).

\section{Additional Numerical Simulations}
\label{sec:num}

We now present a couple of simulations, which intend to provide
a wider perspective of the system behavior.
They
were carried out with
$N=4\times10^3$ units and $\epsilon=1$.

\subsection{The saddle solution}
\label{sec:saddle}

\begin{figure}[b]
    \centering
    \includegraphics[width=0.8\linewidth]{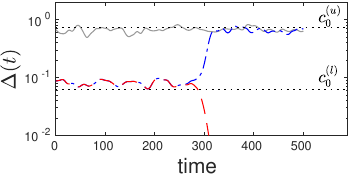}
    \caption{Time evolution of $\Delta(t)$ for a network of size $N=4\times10^3$ with $\epsilon=1$ and $g=0.92$.
The solid grey line is the result of the unperturbed dynamics.
The dashed and dash-dotted lines
correspond to initial conditions obtained rescaling the variance $\Delta$ at $t=0$
by a factor $f=0.228814\ldots$. The factors $f$ differ in $10^{-15}$. After a transient near the saddle,
the trajectory depicted {by a dash-dotted} blue {line} returns
to the chaotic attractor, while the {dashed} red one converges to the trivial fixed point ($\Delta=0$).
Two horizontal dotted lines mark the average variance $c_0$ of
two self-consistent chaotic solutions according to dynamic mean-field theory, see
Figs.~\ref{fig:fold} or \ref{fig:N4000}(b).
    }
\label{fig:saddle}
\end{figure}

We now corroborate the existence
of a saddle solution between the chaotic attractor and the
stable trivial rest state. Adopting $g=0.92$, we let the system evolve in the chaotic attractor
up to at a particular time ($t=0$). At this moment the coordinates of all units
are rescaled by a certain factor $\sqrt{f}$ with respect to the population average.
As we may see in Fig.~\ref{fig:saddle},
at the critical $f$ value
(for our particular numerical experiment $f_c=0.228814\ldots$)
the dynamics initially converges to a saddle state. Remarkably, the
range of values exhibited by the variance $\Delta(t)$ is near the value
$c_0^{(l)}$ predicted in the thermodynamic limit.
For $f$ slightly above $f_c$ the system returns to the chaotic
attractor, while just below the system becomes attracted by the trivial rest state FP$_0$.
This means that the stable manifold of the saddle is the boundary
between the basins of attraction of both attractors.

\subsection{High-dimensional chaos}

\begin{figure}
    \centering
    \includegraphics[width=0.8\linewidth]{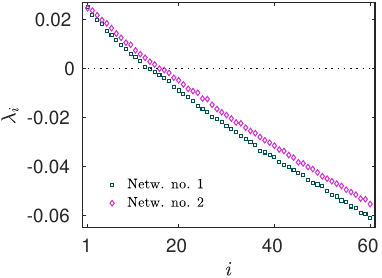}
    \caption{ Sixty largest Lyapunov exponents $\{\lambda_i\}_{i=1,\ldots,60}$
    for the two networks
    in Fig.~\ref{fig:N4000} at $g=1$ ($\epsilon=1$, $N=4\times10^3$).
    The integer part of the Kaplan-Yorke dimension is $28$
    for network {no.}~$1$, and $32$ for network {no.~}$2$.
    }
\label{fig:spectum}
\end{figure}

The standard SCS model exhibits hyperchaos, i.e.~more than one positive Lyapunov exponent
\cite{engelken}.
We have computed the largest Lyapunov exponents
for the two networks in ~Fig.~\ref{fig:N4000},
obtaining $13$ and $15$ positive Lyapunov exponents
for networks {no.}~$1$ and {no.}~$2$, respectively.
The relevant part of the Lyapunov spectrum is shown
in Fig.~\ref{fig:spectum}.

The quasi-continuous appearance of the Lyapunov spectrum suggests
that ---as in the standard SCS model \cite{engelken}--- chaos is extensive,
i.e.~typical measures of chaos (the fractal dimension, the Kolmogorov-Sinai entropy, etc.)
grow linearly with the network size.

\section{Conclusions}

In this work we have analyzed the
classical SCS random neural network
considering
a monoparametric family of odd-symmetric transfer functions $\phi$, which
includes
the usual hyperbolic tangent
as a particular case ($\epsilon=0$).
The third derivative of the transfer
function at zero determines
the abruptness of the edge of chaos.
This might be relevant for tuning the learning capability of neural networks, and
certainly deserves further study.
When the transfer function $\phi(x)$ is concave for all $x>0$ ($\epsilon<1/3$)
the usual continuous transition from the trivial
fixed point to the chaotic phase is observed.
At $\epsilon=1/3$ ($\phi'''(0)=0$)
the transition to chaos becomes particularly abrupt,
while remaining continuous.
Finally, for $\epsilon>1/3$, the transition becomes discontinuous.
{(Retrospectively, this situation had been
already anticipated for the time-discrete SCS model with an
``accelerating nonlinearity'' \cite{toyoizumi11}.)}
{In the discontinuous transition}, the appearance of a pair of heterogeneous fixed points
precedes chaos, which appears out of the blue
in an ``attractor-repeller fold''.
It must be stressed that in low-dimensional
systems such a fold for chaotic sets is not
observed (it may be fabricated, of course, but it is completely non-generic).

Here we have focused on the chaotic and the resting states.
Self-consistent periodic solutions corresponding to periodic
$c(\tau)$ also exist, although they are expected to be unstable.
Unstable periodic orbits constitute the
skeleton of finite dimensional chaotic attractors.
Do they play any relevant role?
A quick analysis indicates that periodic $c(\tau)$ solutions
also appear via fold bifurcations.
However, those with positive energy, able of visiting negative values of $c$,
only appear for $g>g_{\mathrm{ch}}^F$. Is this a generic property of
infinite-dimensional chaos in random neural networks with global $h_i\to-h_i$ symmetry?

In the mindset of comparing with low-dimensional systems, we
found the phase diagram in Fig.~\ref{fig:pd} quite suggestive. The degenerate
point at $(g,\epsilon)=(1,1/3)$ ressembles codimension-two points
encountered in bifurcation theory. The spectral properties
of infinite-dimensional random networks do not permit
usual center manifold reduction. Still, one may wonder
if there exists a systematic classification of bifurcations
for these systems.
At this stage, we can only conclude that the dynamics of
random neural networks represents a challenge for chaos theory.

Finally, it deserves to be emphasized that
this work deals with a network of idealized ``rate neurons''.
In the spirit of \cite{kadmon15} (see also \cite{harish15,angulo17}),
it would be interesting to verify if
the results can be extended to spiking neuronal circuits.

\section*{Acknowledgments}
{I thank Taro Toyoizumi for an enlightening discussion.}
I acknowledge support by
Grant No.~PID2021-125543NB-I00,
funded by MCIN/AEI/10.13039/501100011033 and by ERDF A way of making Europe.

\setcounter{equation}{0}
\renewcommand{\theequation}{A\arabic{equation}}

\section*{Appendix: Heterogeneous fixed point}

\subsection*{Static mean-field-theory}

Besides the homogeneous state FP$_0$ ($h_i=0$), heterogeneous
fixed points exist for large enough $g$.
They are solutions of the fixed-point equation $h_i^*=g\sum_j J_{ij} \phi(h_j^*)$.
In the following calculations
we
assume
the system is self-averaging, see e.g.~the Appendix in
Ref.~\cite{mastrogiuseppe} for a more rigorous treatment.

In the thermodynamic limit, all the heterogeneous solutions
become the same density $\rho(h^*)$.
For large $N$, it is reasonable to expect $J_{ij}$ and $h_j^*$ to become uncorrelated.
Hence, by virtue of the central limit theorem, $\rho(h^*)$ is a normal distribution
around zero, with variance $c_*$.
Self-consistency implies the variance necessarily satisfies
\begin{equation}
c_*=g_{\mathrm{fp}}^2\int_{-\infty}^\infty \phi\left(\sqrt{c_*}z\right)^2 Dz .
\label{selfc}
\end{equation}
From this expression we obtained the parametric curves
$g_{\mathrm{fp}}(c_*)$ in Fig.~\ref{fig:fold}.
(Other order parameters, besides the
variance $c_*$, can be of interest as well \cite{qiu}.)

Taylor expanding Eq.~\eqref{selfc}, one finds
how the heterogeneous solution branches off the trivial fixed point.
We get:
\begin{equation}
c_*=g_{\mathrm{fp}}^2\int_{-\infty}^\infty
\left[\frac{c_*\,  z^2}{2} (\phi^2)''+ \frac{c_*^2 z^4}{24} (\phi^2)'''' + O(c_*^3)\right] Dz ,
\end{equation}
where the derivatives of $\phi^2$ are evaluated at 0;
we have used the identities $\phi(0)=\phi''(0)=0$.
After a few manipulations we obtain the asymptotic dependence of $g_{\mathrm{fp}}$ on $c_*$:
\begin{equation}
g_{\mathrm{fp}}^2(c_*)\simeq\frac1{1+\phi'''(0) c_*+ O(c_*^2)} \simeq 1- \phi'''(0) c_* .
\label{g2}
\end{equation}
where we have assumed $\phi'(0)=1$. It is manifest in Eq.~\eqref{g2} that
the sign of $\phi'''(0)$ determines the orientation
of the branch emanating from $g=1$ in the bifurcation diagram:
If $\phi'''(0)<0$ (i.e., $\phi'(0)$ is a local maximum), then  the usual scenario is
recovered. Instead, if $\phi'''(0)>0$, then the nontrivial fixed point emanates
from FP$_0$ towards $g<1$.
As already written in the main text $\phi'''(0)=-2+6\epsilon$,
and the critical $\epsilon$ value is hence $\epsilon_c=1/3$.

\subsection*{Stability of the heterogeneous fixed point}

As mentioned in the main text, the heterogeneous rest states are always 
found to be unstable.
The linear equation for infinitesimal perturbations is
\begin{equation}
\dot{\delta h}_i=-\delta h_i + g \sum_j J_{ij} \phi'(h_j^*) \delta h_j .
\end{equation}
The Jacobian is minus the identity matrix plus a
inhomogeneous random matrix, with
the elements in the $j$-th column possessing a specific variance $\phi'(h_j^*)^2/N$.
In the thermodynamic limit the eigenvalues of the Jacobian
are contained in a circle of radius squared $r^2=\overline{\phi'(h_j^*)^2}$
centered at $-1$ \cite{rajan06,harish15,aljadeff15}. The eigenvalue with the largest real part is
\begin{equation}
\lambda_{\rm max}= -1 + g \, r ,
\end{equation}
where $r$ satisfies:
\begin{equation}
 r^2=\int_{-\infty}^\infty \left[\phi'\left(\sqrt{c_*}z\right)\right]^2 Dz .
 \label{r2}
\end{equation}
Numerical solution of this system yields positive $\lambda_{\rm max}$ in all cases
we have investigated.
In the small $c_*$ region we need to Taylor expand
the right-hand side of Eqs.~\eqref{selfc} and \eqref{r2} at order
$c_*^3$ and $c_*^2$, respectively.
After some algebra, we get the asymptotic value of the dominating eigenvalue
for the rest state branching off FP$_0$:
\begin{equation}
\lambda_{\rm max}=6\left(\epsilon-\frac13\right)^2 c_*^2 +O(c_*^3) .
\end{equation}
Remarkably, $\lambda_{\rm max}$ remains positive for all $\epsilon$
(save for $\epsilon=1/3$, in which case the next order in $c_*$ should be computed).

\end{document}